\documentclass[12pt]{article}

\usepackage[dvips]{graphicx}
\begin{document}

\title{The impact of thermal noise on kink propagation through a heterogeneous system }

\author{
{\sc  J. Gatlik and T. Dobrowolski}
\\ Institute of Physics UP,
\\ Podchor\c{a}\.zych  2, 30-084 Cracow, Poland}

\maketitle

\begin{abstract}
The impact of thermal noise on kink motion through the curved
region of the long Josephson junction is studied. On the basis of
the Fokker-Planck equation the analytical formula that describes
the probability of transmission of the kink over the potential
barrier is proposed. The analytical results are compared with the
simulations based on the field model.
\end{abstract}

\eject

\section{Introduction}

In recent years, a significant increase of interest in the
construction of a variety of appliances that use superconducting
elements is observed. Among the devices manufactured on the basis
of superconductors, Josephson junctions occupy a prominent
position. The effect of supercurrent flow without any voltage
applied was initially predicted by Brian D. Josephson
\cite{Josephson1962,Josephson1974}. A device known as a Josephson
junction consists of two superconductors coupled by some weak
link. The weak link can be made of a thin insulating barrier (in
S-I-S junctions), normal non-superconducting metal (in S-N-S
junctions), or have a form of constriction that weakens the
superconductivity at the point of contact (in S-s-S junctions).
The effect was experimentally confirmed by Philip Anderson and
John Rowell \cite{Anderson1963}.

Presently there are a variety of devices which contain Josephson
junctions in their design \cite{Seidel2015,Braginski2019}. They
can be classified into three groups. In the first group one can
include antennas, amplifiers, filters, bolometers, single photon
detectors, magnetometers and many others. The second group
consists of digital electronic appliances like digital-to-analogue
and analogue-to-digital converters and rapid single flux quantum
computing elements. The third group consists of quantum computing
devices.

In the context of future technical applications of the Josephson
junction it is natural to look for superconducting materials with
high critical temperatures. Presently at normal pressure, it is
possible to achieve a state of superconductivity at relatively
high temperatures in the so-called high-temperature
superconductors \cite{Bednorz1986}. An example of such materials
is cuprate-perovskite ceramic which has a critical temperature
above 90 K. Nowadays one of the highest - temperature
superconductors is $HgBa_2Ca_2Cu_3O_{8+\delta}$ with a critical
temperature exceeding 133 K \cite{Schilling1993}. In particular,
exceeding the temperature  77 K allows the use of liquid nitrogen,
on an industrial scale, in cooling systems of superconducting
devices.

The Josephson junction properties required for optimal performance
of the appropriate devices can be planned at the design stage of
the equipment that uses them. Between multiple approaches directed
at obtaining requested properties of Josephson junctions, shape
engineering plays a significant role. In this approach, particular
modifications of the junction shape are proposed in order to
obtain their required properties. For example, in the article
\cite{Benabdallah1996} the authors proposed a device that consists
of a junction with an exponentially tapered width, decreasing
toward the load. In this device the junction is preceded by an
idle region, where the oxide layer is thicker, preventing the
tunnelling of Cooper pairs.

On the other hand, in the heart-shaped annular junction two
classical vortex states can be prepared, corresponding to two
minima of the potential \cite{Kemp2002a}. The bias current across
the junction is used to slant the potential. The strength and
direction of the the applied external magnetic field plays the
role of the control parameters. For example, all these parameters
can be used in order to modify the barrier height. The
heart-shaped long Josephson junction placed in an in-plane
external magnetic field was also considered in article
\cite{Kemp2002b}. Based on this geometry the authors designed a
classical system with two ground states. At sufficiently low
temperatures, this structure is expected to behave as a quantum
two-state system.

The other opportunity to modify properties of a junction is
formation of the T-shaped geometry \cite{Gulevich2006a}. The above
mentioned appliance consists of two perpendicular Josephson
T-Lines forming a T junction. The particular effect present in the
device is the creation of a new vortex when an original vortex,
moving along the main Josephson T-line, is passing the T junction.
The new vortex created at the T junction begins its motion in the
direction perpendicular to the main Josephson T-line. The creation
of a new vortex is substantially dependent on the energy of the
original vortex. If the kinetic energy of the original vortex is
too small then the T junction acts as a barrier and the original
vortex is reflected without creation of a new vortex.

A similar, to some degree, proposal is sigma-pump. The main
advantage of this system is the lack of the barrier associated
with the T-junction present in the T-pump. Instead, the Josephson
transmission line is connected with the ring smoothly through the
Y junction. In this pump a nucleation barrier is absent. Moreover,
the nucleation energy is gathered by the trapped fluxon during its
motion in the potential associated with increasing width. A
similar system is considered in articles \cite{Gulevich2008} and
\cite{Caputo2014}.

An interesting possibility is an annular junction delimited by two
closely spaced confocal ellipses that is characterized by a
periodically modulated width \cite{Monaco2016,Monaco2018}. This
spatial dependence, in turn, produces a periodic potential that
interchangeably attracts and repels the fluxons. In this
particular junction double-well potential, experienced by an
individual fluxon, is produced by an intrinsic non-uniform width.

If the thickness of the dielectric layer in the junction is
position dependent then the kink experiences the effective
potential originating in heterogeneities present in the system
\cite{Dobrowolski2020}. The value of transmission critical current
in this case is strictly determined by the parameters of the
system.

The effects of arbitrary curvature on fluxon motion in curved
Josephson junctions were studied in articles
\cite{Dobrowolski2009,Dobrowolski2011,Dobrowolski2012,Dobrowolski2013,Dobrowolski2017}
with curvatures playing the role of potential barriers for kink
motion. In particular in \cite{Gatlik2021} the different
simplified effective descriptions were compared in order to choose
the most suitable for the considered system.

In the present article we study the curved system with bias
current and the quasi-particle dissipation taken into account.
Moreover this paper is aimed at studying the effects of nonzero
temperature of the system on the process of penetration of the
potential barrier through the kink. We present the appropriate
analytical results and compare them with the results of
simulations performed in the field model. The analytical
approximations rely mainly on the projection onto energy density
method.

\section{Kink in curved system}
We consider the kink motion in the sine-Gordon model with position
dependent dispersive term
\begin{equation}\label{1}
  \partial_t^2 \phi + \alpha  \partial_t \phi - \partial_x ({\cal F}(x) \partial_x
  \phi ) + \sin \phi = - \Gamma .
\end{equation}
The physical motivation for description of curvature effects in
the framework of this model was described in detail in the
articles \cite{Gatlik2021,Dobrowolski2012}. In this paper it was
shown that this modification appears in the description of a
curved Josephson junction. The function ${\cal F}(x)$ contains
information about curvature of the junction. The kink solution in
this physical situation represents the fluxon propagating along
the long junction. In the context of the Josephson junction the
distances in the above equation are measured in the units of
Josephson penetration depth, the time is measured in units of the
inverse plasma frequency, $\alpha$ represents the dissipation
caused by the quasi-particle currents and $\Gamma$ is bias
current. Reduction of the field model to a single mechanical
degree of freedom is performed in the framework described in
article \cite{Gatlik2021} procedure called projection onto energy
density. In order to realize this scheme we introduce into field
equation the kink like ansatz
$$ \phi(t,x) = 4 \arctan (e^{\xi(t,x)}),$$
where in nonrelativistic limit the function $\xi(t,x)$ is
approximated as follows
$$\xi = x - x_0(t). $$
Here $x_0(t)$ denotes a position of the kink. For further
convenience we introduce auxiliary function $g(x)$
$$ {\cal F}(x) = 1 + \varepsilon g(x),$$
where $\varepsilon$ is a dimensionless parameter that controls the
magnitude of heterogeneity. We consider the deformation of the
system localized between $x=x_i$ and $x=x_f$. To be precise we
assume the function $g(x)$ in the form
$$ g(x) = \theta(x-x_i) - \theta(x-x_f).$$
In the context of the curved junction the form of this function
means constant (nonzero) curvature located between $x_i$ and
$x_f$. In these markings, the field equation (1) can be
transformed to the form
\begin{eqnarray}\label{2}
\dot{u} \textrm{~sech} \xi + u^2 \textrm{~sech} \xi \tanh \xi +
 \alpha\, u \textrm{~sech} \xi  =    \\  - \varepsilon\, \partial_x g(x)
    \textrm{~sech} \xi + \varepsilon\, g(x) \textrm{~sech} \xi \tanh
    \xi + \frac{1}{2} \,\Gamma ,     \nonumber
\end{eqnarray}
where $u$ denotes the kink speed i.e. $u \equiv \dot{x}_0.$ The
projection onto energy density in co-moving, with kink, reference
frame relies on integration of the equation (2) with the energy
density profile
\begin{equation}\label{3}
 Eq=0 \Rightarrow \int_{- \infty}^{+ \infty} dx \textrm{~sech}^2 \xi \, Eq
 = 0 .
\end{equation}
This procedure is quite similar to the projection onto zero mode
of the kink. The only difference lies in the fact that this
profile is better localized in the neighborhood of the kink
position. There is also a more fundamental reason for choosing
this profile namely in systems with explicitly broken invariance
with respect to spatial translations the zero mode, strictly
speaking, does not exist while the energy density is still well
defined. As the final outcome of elimination of the space variable
we obtain the equation for the kink position
\begin{equation}\label{4}
  \dot{u} + \alpha u = \frac{2}{\pi}\, \Gamma - \varepsilon \frac{4}{3
  \pi} \left( \textrm{~sech}^3(x_i -x_0(t)) - \textrm{~sech}^3(x_f
  -x_0(t))\right) ,
\end{equation}
where during integration we used the formula
$$ \partial_x g(x) = \delta(x-x_i) - \delta(x-x_f).$$
The bracket from the right side of the equation (4) represents the
force originated in the curved region of the junction.
\begin{figure}[htbp]
\centering
\includegraphics[width=10cm]{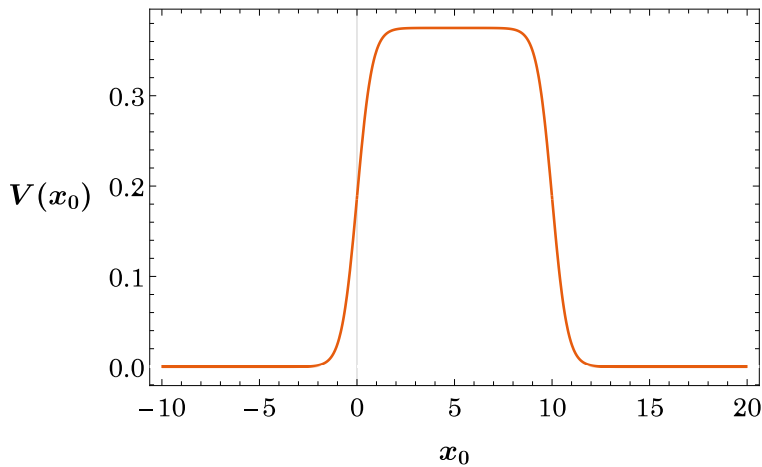}
\caption{The potential $V(x_0)$ that represents the presence of
the curved region located between $x_i=0$ and $x_f=L=10$. The
parameter $\varepsilon$ is equal to one.}
\end{figure}
The potential for this force represents the barrier associated
with the curved region (Fig.1)
\begin{eqnarray}\label{5}
V(x_0)= \varepsilon \frac{3}{4 \pi} \, \Large[
\arctan(\tanh(\frac{x_0-x_i}{2})) -
\arctan(\tanh(\frac{x_0-x_f}{2})) +    \\  \frac{1}{2} \, {\rm
sech}(x_0 - x_i) \tanh(x_0 - x_i) - \frac{1}{2} \, {\rm sech}(x_0
- x_f) \tanh(x_0 - x_f) ] .    \nonumber
\end{eqnarray}
In this paper location of the inhomogeneity is assumed to be
between $x_i=0$ and $x_f=L$.

In order to estimate the value of the critical speed that
separates the kinks reflected from the barrier from those which
pass over the barrier we separate the problem of movement in the
barrier potential from the motion under the influence of constant
force represented by constant bias current. The total energy of
the kink that moves in the potential $V(x_0)$ is $E=\frac{1}{2}
m_0 {u}^2+ V(x_0),$ where kink mass is equal $m_0=8.$ At the
beginning of its motion the kink moves almost freely having only
kinetic energy
$$E_{in} = \frac{1}{2}
m_0 {u_c}^2 .$$ We assume that at the end of its motion the kink
stops on the top of the barrier having only the potential energy
$$E_{fin} = V(x_0=L/2) = \frac{32}{3 \pi}~ \varepsilon \left[ 2
\arctan \left( \tanh \frac{L}{4}\right) + \mathrm{sech}\frac{L}{2}
\tanh \frac{L}{2}\right] .$$ The conservation of the energy leads
to the following estimation of the critical velocity
\begin{equation}\label{6}
  u_c=\sqrt{\frac{8}{3 \pi}~ \varepsilon~} ~\sqrt{ 2 \arctan \left( \tanh \frac{L}{4}\right)+ \mathrm{sech}\frac{L}{2} \tanh \frac{L}{2}
  } .
\end{equation}
This estimation quite well describes the values of the critical
velocity even in the case when the bias current and the
dissipation term are taken into account. The reason for this is
the fact that we work with velocities for which the bias current
and dissipation almost cancel each other.

\section{The influence of thermal fluctuations on the kink motion}
Far from the barrier the last bracket from the right hand side of
the equation (4) describes the residual interaction of the kink
with the barrier (which is a consequence of the interaction of the
kink tail with the curved region). We will describe how the fluxon
approaching the barrier from the left, interacts with this
barrier. This residual impact will be treated approximately as a
position independent small interaction and therefore we consider
the following equation
\begin{equation}\label{7}
  \dot{u} + \alpha u = \frac{2}{\pi}\, \Gamma - r ,
\end{equation}
instead of equation (4). Here $r$ is the above mentioned small
residual interaction.

Our intention is to describe the influence of the nonzero
temperature, of the system, on the process of overcoming the
barrier by the fluxon. We assume that the bias current is a random
variable i.e. it fluctuates due to the non-zero temperature of the
system. The average value of the bias current we denote by
$\Gamma_0$
\begin{equation}\label{8}
   <\Gamma(t)> = \Gamma_0 ,
\end{equation}
where averaging is with respect to all realizations of the thermal
noise. In this situation from equation (7) we calculate the
average value of the stationary speed (in stationary case
$<\dot{u}>=0$)
\begin{equation}\label{9}
 u_s  = \frac{2  }{\pi \alpha} \, \Gamma_0 - \frac{r}{\alpha} ,
\end{equation}
where the average value of the stationary velocity is denoted by
$u_s$. Moreover the thermal noise has the character of a white
Gaussian noise and therefore the time correlation function of the
bias current we assume in the form
\begin{equation}\label{10}
<\Gamma(t) \Gamma(t') > = A \delta(t-t') .
\end{equation}
In order to fix the appropriate value of the prefactor $A$ for the
system in thermal equilibrium we came back to the equation (7).
The solution of this equation under assumption of constant $r$
reads
\begin{equation}\label{11}
u(t) = \frac{2}{\pi} \int_0^t dt'\, \Gamma(t') e^{\alpha (t'-t)} -
\frac{r}{\alpha} \, (1 - e^{-\alpha t}) .
\end{equation}
Now we are ready to calculate the time correlation function of the
velocity
\begin{eqnarray}\label{12}
& < u(t)u(\bar{t}) > = \left( \frac{2}{\pi}\right)^2 \int_0^t dt'
\int_0^{\bar{t}} dt'' <\Gamma(t') \Gamma(t'')>
e^{\alpha(t'+t''-t-\bar{t})} &  \\
& - \frac{r}{\alpha}\, (1-e^{-\alpha \bar{t}}) \frac{2}{\pi}
\int_0^{t} dt' <\Gamma(t')> e^{\alpha (t'-t)} - \frac{r}{\alpha}\,
(1-e^{-\alpha t}) \frac{2}{\pi} \int_0^{\bar{t}} dt' <\Gamma(t')>
e^{\alpha (t'-\bar{t})} & \nonumber \\  & + \left(\frac{r}{\alpha}
\right)^2 (1-e^{-\alpha \bar{t}}) (1-e^{-\alpha t}) & \nonumber
\end{eqnarray}
If we apply formulas (8) and (10) for average and the time
correlation of the bias current, and moreover assume $t=\bar{t}$
then we obtain
\begin{equation}\label{13}
  <u(t)^2> =<u(t) u(\bar{t})>_{\bar{t}=t} = \frac{2 A}{\pi^2
  \alpha}\, (1-e^{-2 \alpha t}) +\left[\left( \frac{r}{\alpha}\right)^2 - \frac{4 \Gamma_0}{\pi \alpha^2} \,r
  \right] (1 - e^{- \alpha t})^2 .
\end{equation}
The system after the required length of time tends to
thermodynamic equilibrium and therefore we extract in the last
formula the terms that dominate long time behaviour of $<u^2>$
\begin{equation}\label{14}
  <u(t)^2> = \frac{2 A}{\pi^2
  \alpha} +\left( \frac{r}{\alpha}\right)^2 - \frac{4 \Gamma_0}{\pi \alpha^2}
  \,r .
\end{equation}
The kinetic energy of the fluxon after a sufficiently long time
reads
\begin{equation}\label{15}
 E_k = \frac{1}{2}\, m <u(t)^2> = \frac{1}{2}\, m \left[\frac{2 A}{\pi^2
  \alpha} +\left( \frac{r}{\alpha}\right)^2 - \frac{4 \Gamma_0}{\pi \alpha^2}
  \,r \right] ,
\end{equation}
where $m$ is kink mass. We expect that after an appropriately long
time the system tends to thermal equilibrium. On the other hand in
thermodynamic equilibrium, on the basis of the equipartition
principle, it is proportional to the temperature $T$
\begin{equation}\label{16}
 E_k = \frac{1}{2}\, k T ,
\end{equation}
here $k$ is Boltzmann constant.  Comparison of the equations (15)
and (16) allows the determination of the coefficient $A$
\begin{equation}\label{17}
A = \frac{\pi^2 \alpha k (T - \Delta T)}{2 m},
\end{equation}
where we denoted
\begin{equation}\label{18}
 \Delta T \equiv \frac{m}{k} \left[ \left( \frac{r}{\alpha}\right)^2 - \frac{4 \Gamma_0}{\pi \alpha^2} \, r
 \right] .
\end{equation}
For further convenience, we transform the formula (18) to the form
containing the critical value of the bias current $\Gamma_c$
\begin{equation}\label{19}
 \Delta T = \Omega (\Gamma_c - \Gamma_0) - \omega .
\end{equation}
This critical value  $\Gamma_c$ separates the values of the bias
current for which the particle passes over the barrier from the
values for which the reflection occurs. The parameters in the
above formula are defined as follows
$$ \omega \equiv \frac{m}{k} \left[  \frac{4 \Gamma_c}{\pi \alpha^2} \, r - \left( \frac{r}{\alpha}\right)^2 \right], \,\,\,\,\Omega \equiv \frac{4 m}{\pi \alpha^2 k}.$$
Finally, the average and the time correlation function of bias
currents are defined by the formulas
\begin{equation}\label{20}
<\Gamma(t)> = \Gamma_0, \,\,\, <\Gamma(t) \Gamma(t')> =
\frac{\pi^2 \alpha k (T - \Delta T)}{2 m} \, \delta(t-t') .
\end{equation}
The equations (20) are the starting point for the derivation of
the Fokker - Planck equation described in appendix A. The
stationary solution of this equation is the following
\begin{equation}\label{21}
   P(u) = \sqrt{\frac{m}{2 \pi k (T-\Delta T)}} \exp\left(-\frac{m}{2 k (T - \Delta T)}\, (u - u_s)^2 \right)
   .
\end{equation}
This probability is a base for calculation of the total
probability of the transmission of the kink through the potential
barrier. We have to deal with the transition event whenever the
kink speed exceeds the critical velocity
\begin{equation}\label{22}
  \Delta P = \int_{u_c}^{\infty} d u P(u) = \frac{1}{2} \, \mathrm{ erf}
  \left( \sqrt{\frac{m}{2 k (T - \Delta T)}} \mid u_c - u_s \mid
  \right) ,
\end{equation}
in this formula $\mathrm{erf}$ denotes an error function. This
probability depends, in addition to temperature and residual
effects, on the difference of critical $u_c$ and stationary $u_s$
velocities in the system. The critical velocity separates two
regimes. In the first regime the particle passes over the barrier
and in the second it reflects from the barrier.

The probabilities obtained on the basis of the field model and
analytical result (22) based on the Foccker - Planck approach are
compared in Figures 3-5 for different ranges of temperatures. Due
to potential applications also in high-temperature
superconductors, the comparison was made for intervals from zero
to  $T = 50 K$, $T = 20 K$ and $T = 5 K$. In all plots the
parameters of the shift $\Delta T$ given in the formula (19) are
fitted so that they take the values $\Omega=25220.6$ and
$\omega=-0.398529$. We decided to fit these parameters because the
residual interaction is out of our control. Starting from the
field model we obtain the fit of $\Delta T$ as a function of the
absolute value of separation between actual average value of the
bias current and its critical value. This fit is presented in
Figure 2.
\begin{figure}[htbp]
\centering
\includegraphics[width=10cm]{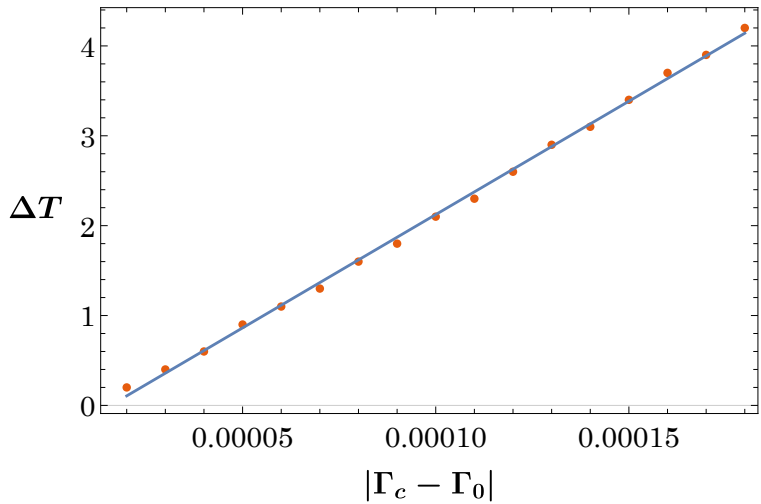}
\caption{ $\Delta T$ as a function of modulus of difference of the
critical value of the bias current and actual average of the bias
current. The parameters of the fit are $\Omega=25220.6$ and
$\omega=-0.398529$. }
\end{figure}
In all simulations we assume damping coefficient on the level of
$\alpha=0.01$. Moreover we assumed the size of the inhomogeneity
$L=10$ and we located its position between $x_i=0$ and $x_f=10$.
The the strength of the heterogeneity is fixed at the level of
$\varepsilon=1$.  Because a relativistic formula (44) for
stationary speed is known therefore we use it in all plots (see
Appendix C). On the other hand in Figures 3-4 the critical
velocity is approximated by the non-relativistic formula (6).
\begin{figure}[htbp]
\centering
\includegraphics[width=10cm]{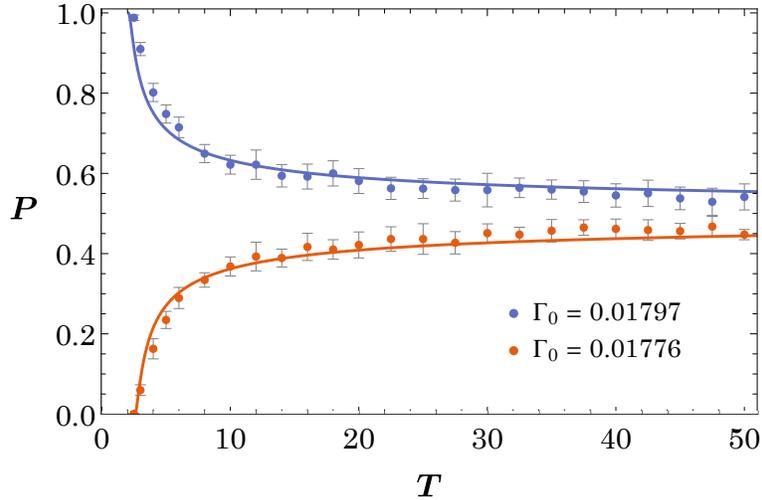}
\caption{ The probability of transition of the fluxon obtained
from the field model compared in the interval $T \in [0 K,50 K]$
with the analytical formula. The parameters of the plots are
$\alpha=0.01$, $\varepsilon=1$, $L=10$. The blue line and points
correspond to the bias current exceeding its critical value and
the red to the bias current below its critical value. }
\end{figure}
\begin{figure}[htbp]
\centering
\includegraphics[width=10cm]{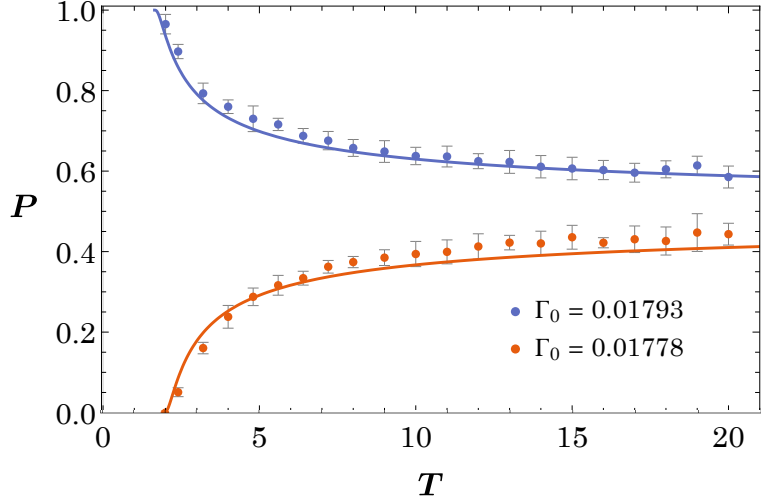}
\caption{ The probability of transition of the fluxon in the
interval $T \in [0K,20K]$. Comparison of the analytical result
with the field model prediction. The parameters of the plots are
$\alpha=0.01$, $\varepsilon=1$, $L=10$. The blue line and points
represents data for the bias current exceeding its critical value
and the red ones the bias current below its critical value. }
\end{figure}
This choice is motivated by the fairly good compatibility of the
approximated formula with the results obtained on the background
of the field model. On the other hand in the case of Figure 5 the
accuracy of the formula (6) was insufficient and therefore we used
the relativistic model (40) obtained in appendix B.
\begin{figure}[htbp]
\centering
\includegraphics[width=10cm]{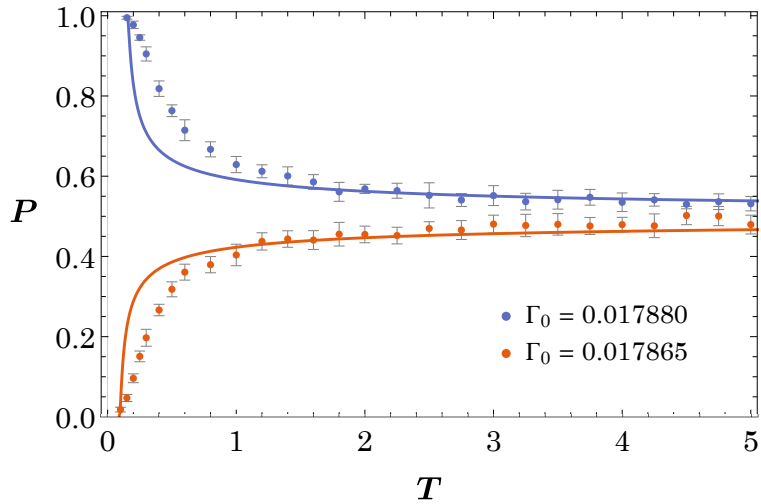}
\caption{Transition probability of the fluxon in the interval $T
\in [0K,5K]$.  The parameters of the plots are $\alpha=0.01$,
$\varepsilon=1$, $L=10$. The blue line and points correspond to
the bias current exceeding its critical value and the red ones to
the bias current below its critical value.  }
\end{figure}
The figures show that for the bias currents above the critical
value (blue line for analytical formula and points for the field
model), as the temperature increases, the probability of the
particle passing over the barrier decreases. The reduction of
transition probability is in the direction of the value of
one-half, the achievement of which would make such a process
completely random. On the other hand, for the bias currents below
its critical value (red line for the formula and points for the
field model) as the temperature increases, the probability of the
particle passing over the barrier also increases. The probability
increases gradually towards the half value beyond which the
process would be completely random. The comparison of the results
of the field model in nonzero temperature with analytical
description provided by formula (22) shows a pretty good level of
compatibility. The results are consistent in Figures 3 and 4,
while in Figure 5 there are deviations below one Kelvin. Figures
3-5 show simulations when the currents slightly differ from the
critical current. On the other hand, if the difference between the
average bias current and its critical value is significant, then
thermal fluctuations have a negligible impact on the process of
interaction between the kink and the curved region. In this case
the interaction is properly described by a deterministic model
i.e. if the bias current is below the value of critical current
then the kink is reflected from the curved region. On the other
hand if bias current exceeds its critical value then the kink goes
through the curved region. This behaviour is crucial for possible
technical applications.

\section{Remarks}
In the present article we considered the impact of  thermal
fluctuations on the process of interaction of the kink with the
heterogeneous region of the system described by a nearly
integrable sine-Gordon model. The physical background of the
studies is the influence of the curvature on the fluxon motion in
the long Josephson junction. We obtained analytical formulas that
describe probabilities of transition through and reflection of the
kink from the potential barrier that represents heterogeneity. The
main result is based on the Foccker - Planck equation obtained for
the considered system. We compared the analytical results with the
simulations performed in the framework of the field model for
different ranges of temperatures. Due to potential applications in
normal and also in high-temperature superconductors, the
comparison was made for intervals from zero to  $T = 50 K$, $T =
20 K$ and $T = 5 K$. The compatibility of the analytical formula
with the numerical simulations is satisfactory in the first
(Fig.3) and the second regime (Fig.4). In the third regime (Fig.5)
the compliance above one $1 K$ is also satisfactory.

The most problematic regime of temperature is presented in figure
5. In this interval we resigned from the formula (6) for critical
velocity and in order to obtain a better fit we used the
relativistic model (40) for estimation of the critical speed.
Either way we observed in the small temperature regime presented
in Figure 6 some discrepancy between the result of the field model
and our fit located in the interval from $0 K$ to $1 K$. We
identified a probable reason for this problem.

In the low temperature regime we observed occurrence of the
resonance windows in the transition process. It means that we
observe very narrow regimes of the parameters that corresond to
transition below the critical speed and moreover the reflection
regimes above the critical velocity. This phenomenon has a place
in the effective model (40) and in the original field model (1) as
well. This phenomenon is responsible for the ambiguity of the
estimation of the critical speed and is responsible for the
discrepancy of the approximate description and the results of the
field model in Fig.6. A similar phenomenon was previously observed
by many researchers. For example in article \cite{Fei1992} in the
$\phi^4$ model an interaction of the kink with attracting point
impurity was studied. The existence of resonance windows in
initial speeds below some threshold velocity had found an
explanation in the resonant energy exchange between the kink
internal mode and its translational mode. This behaviour was first
observed numerically by Campbell \cite{Campbell1983} and his
collaborators in the case of kink-antikink scattering in the
$\phi^4$ model. Presently there is a variety of articles that
contain a detailed explanation of the two-bounce resonance
observed in kink - antikink collisions \cite{Goodman2005}. A
separatrix map for this problem that explains the complex
fractal-like dependence on initial velocity for kink-antikink
collisions was also constructed. The chaotic nature of such
collisions depends on the transfer of energy to a secondary mode
of oscillation \cite{Goodman2008}. In the frame of the moduli
space formalism \cite{Manton2021} a spectacular result in
reproducing the fractal structure in the formation of the final
state was reached in article \cite{Adam2022}. The key insight of
these articles is that the existence of resonance windows is
possible due to the presence of an internal mode in the spectrum
of the kink in the $\phi^4$ model \cite{Takyi2016,Kevrekidis2019}.
The situation in the case of the sine-Gordon model is different.
The linear spectra of the kink excitations does not contain the
discrete internal mode and therefore the structure and the nature
of the windows in the model considered in this paper is enigmatic.
On the other hand the modification of the sine-Gordon model
considered in this article belongs to the so called nearly
integrable variations of the original model. The studies on this
subject are ongoing and will be presented in the future. To some
degree a similar example of the model containing resonance windows
in kink-antikink interactions was presented in the article
\cite{Dorey2011}. This article describes the solutions of the
$\phi^6$ model that does not contain, in its linear spectra of
excitations, the discrete internal eigenmodes which, to some
degree, resembles our system.

\section{Acknowledgements}
This research was supported in part by PLGrid Infrastructure.

\section{Appendix A: Fokker - Planck equation}
For the sake of completeness of the article we present derivation
of the Fokker - Planck equation for the system studied by us.
First, let us notice that velocity variation is a random variable
with the following mean value
\begin{equation}\label{23}
  <\delta u> = <\dot{u} \, \delta t> =(- \alpha u - r + \frac{2}{\pi}\, \Gamma_0 ) \, \delta
  t ,
\end{equation}
where we used formula (7) and bias current mean value (8).
Similarly, formula (20) leads to the expression
\begin{equation}\label{24}
  <\delta u \delta u> = \frac{2 \alpha k (T -\Delta T)}{m} \, \delta
  t .
\end{equation}
Next the conditional probability that the particle which has
velocity $u$ at time $t + \delta t$, a moment earlier i.e. at
$\bar{t}$, had velocity $\bar{u}$ we denote by $P(u,t+\delta t;
\bar{u},\bar{t})$. Taylor expansion of this probability with
respect to final time reads
\begin{equation}\label{25}
   P(u,t+\delta t;
\bar{u},\bar{t}) = P(u,t; \bar{u},\bar{t})+\partial_t P(u,t;
\bar{u},\bar{t}) \delta t ,
\end{equation}
where we ignored the terms of second and higher orders in $\delta
t$. On the other hand we can obtain this expansion starting from
the Chapman - Kolmogorov equation
\begin{equation}\label{26}
   P(u,t+\delta t; \bar{u}, \bar{t} ) = \int_{-\infty}^{+\infty} d
   u'  P(u,t+\delta t; u', t' )  P(u',t'; \bar{u}, \bar{t}),
\end{equation}
which states that, at some intermediate time $\bar{t}<t'<t+\delta
t$ the velocity $u'$ belongs to the interval $u'\in
(-\infty,+\infty)$. The probability present in this formula can be
expressed with the velocity variation $\delta u$ as follows
\begin{equation}\label{27}
 P(u,t+\delta t; u', t' ) =<f(u-u'-\delta u> ,
\end{equation}
which can be expanded with respect to velocity
\begin{equation}\label{28}
 P(u,t+\delta t; u', t' ) =f(u-u') + <\delta u> \partial_{u'} f(u-u') + \frac{1}{2} \,
<\delta u \, \delta u> \partial_{u'}^2 f(u-u') .
\end{equation}
Truncation at the second order is motivated by the fact that they
contain at most linear terms in $\delta t$. The Chapman -
Kolmogorov formula now reads
\begin{eqnarray}\label{29}
& P(u,t+\delta t; \bar{u}, \bar{t} ) = \int_{-\infty}^{+\infty} d
   u'  [ f(u-u') +  & \\  & <\delta u> \partial_{u'} f(u-u') + \frac{1}{2} \,
<\delta u \, \delta u> \partial_{u'}^2 f(u-u') ]  P(u',t';
\bar{u}, \bar{t}),  &   \nonumber
\end{eqnarray}
Assuming that $f$ and its first derivative disappear at plus/minus
infinity and integrating second and third terms by parts we obtain
\begin{equation}\label{30}
   P(u,t+\delta t; \bar{u}, \bar{t} ) = \int_{-\infty}^{+\infty} d
   u'  f(u-u')[ P - \partial_{u'} (<\delta u> P) + \frac{1}{2} \,
   <\delta u \, \delta u> \partial_{u'}^2 P] ,
   \end{equation}
where we have used the fact that $<\delta u\, \delta u>$ does not
depend on $u$. From formulas (23) and (24) it is also transparent
that the last two terms are linear in $\delta t$. Let us also
notice that without random variation $\delta u$ the probability
distribution is unambiguously determined as follows $f(u-u') =
\delta(u-u')$ and therefore after integration we obtain
\begin{eqnarray}\label{31}
&  P(u,t+\delta t; \bar{u}, \bar{t} ) =P(u,t; \bar{u}, \bar{t} ) -
  \partial_u \left( <\delta u> P(u,t; \bar{u}, \bar{t} ) \right) + \\  & \frac{1}{2} \, <\delta u \delta u> \partial_u^2 P(u,t; \bar{u},
\bar{t} ) .  &   \nonumber
\end{eqnarray}
Next we replace the average and variance of the random variable
$\delta u$ from formulas (23), (24) and we obtain
\begin{eqnarray}\label{32}
&  P(u,t+\delta t; \bar{u}, \bar{t} ) =P(u,t; \bar{u}, \bar{t} ) +
  \partial_u \left( (\alpha u + r - \frac{2}{\pi} \, \Gamma_0) P(u,t; \bar{u}, \bar{t} ) \right) \delta t + \\  &   \frac{\alpha k (T-\Delta T)}{m} \, \partial_u^2 P(u,t;
\bar{u}, \bar{t} ) \delta t . &   \nonumber
\end{eqnarray}
Finally, comparison of the above formula with equation (25) leads
to the following form of the Fokker - Planck equation for the
system considered by us
\begin{equation}\label{33}
 \partial_t P =  \partial_u \left( (\alpha u + r - \frac{2}{\pi} \, \Gamma_0) P  +
 \frac{\alpha k (T-\Delta T)}{m} \, \partial_u P \right) .
\end{equation}
The time independent ($\partial_t P=0$) normalized solution of
this equation reads
\begin{equation}\label{34}
   P(u) = \sqrt{\frac{m}{2 \pi k (T-\Delta T)}} \exp\left(-\frac{m}{2 k (T - \Delta T)}\, (u - u_s)^2 \right) ,
\end{equation}
where we used (9) in order to identify the presence of the average
stationary velocity $u_s$ in the equation.

\section{Appendix B: Effective relativistic description of the kink}
In order to obtain a relativistic approximation of the critical
speed of the kink we reconsider the projection procedure onto the
energy density used in Section 2. We start again with the field
equation
\begin{equation}\label{35}
  \partial_t^2 \phi + \alpha  \partial_t \phi - \partial_x ({\cal F}(x) \partial_x
  \phi ) + \sin \phi = - \Gamma_0 .
\end{equation}
Similarly as before we introduce the kink like ansatz into the
field equation
$$ \phi(t,x) = 4 \arctan (e^{\xi(t,x)}),$$
where this time the function $\xi$ takes its relativistic form
$$\xi = \gamma(t)(x - x_0(t)). $$
Moreover, the function ${\cal F}$ is expressed by the auxiliary
function $g(x)$
$$ {\cal F}(x) = 1 + \varepsilon g(x),$$
where dimensionless parameter $\varepsilon$ controls the magnitude
of curvature. Next we insert the kink ansatz into the field
equation (35), obtaining
\begin{eqnarray}\label{36}
&  \left[ \left( \frac{\ddot{\gamma}}{\gamma}+ \alpha
\frac{\dot{\gamma}}{\gamma}\right)\xi - (2 \dot{\gamma} u + \gamma
\dot{u} + \alpha \gamma u) \right] \textrm{~sech} \xi + & \nonumber \\
 &
 \left[ (\gamma^2-1-\dot{\gamma}^2 u^2) + 2 \dot{\gamma}
u \xi - \left( \frac{\dot{\gamma}}{\gamma} \right)^2 \xi^2 \right]
\mathrm{sech} \xi \tanh \xi - &    \\ & \varepsilon \gamma (
\partial_x g )~ \mathrm{sech} \xi + \varepsilon \gamma^2
g(x)~\mathrm{sech} \xi \tanh \xi = - \frac{1}{2} \Gamma_0 .&
\nonumber
\end{eqnarray}
We eliminate the spatial variable from the description by
projection onto the energy density distribution
\begin{equation}\label{37}
 Eq=0 \Rightarrow \int_{- \infty}^{+ \infty} dx \textrm{~sech}^2 \xi \, Eq
 = 0 .
\end{equation}
As a result of this procedure, we obtain a one-dimensional
relativistic model describing the location of the kink
\begin{equation}\label{38}
   \dot{u} + \alpha u + \frac{4}{3} ~u \frac{\dot{\gamma}}{\gamma}
   = \frac{4 }{3 \pi }~\varepsilon \gamma \left( \mathrm{sech}^3 \xi_L - \mathrm{sech}^3 \xi_0
   \right) + \frac{2}{\pi \gamma} ~\Gamma_0 ,
\end{equation}
where we denoted  $\xi_L = \gamma (L-x_0(t))$ and $\xi_0 = \gamma
(-x_0(t)) .$ On the other hand, introducing to the last equation
the Lorentz factor $ \gamma = 1/\sqrt{1 - u^2}$ (we use the units
with Swihart velocity equal to one $c=1$) we obtain
\begin{equation}\label{39}
\left( 1 + \frac{1}{3}~ u^2 \right) \dot{u} + \alpha u (1-u^2)
=\frac{4 }{3 \pi }~\varepsilon \sqrt{1-u^2} \left( \mathrm{sech}^3
\xi_L - \mathrm{sech}^3 \xi_0
   \right) + \frac{2}{\pi } (\sqrt{1-u^2})^3 ~\Gamma_0 .
\end{equation}
This equation is a base for estimation of the critical speed in
our system in the low temperature regime presented in Fig. 5.

\section{Appendix C: Relativistic approximation of the stationary speed}

For the sake of completeness of the presentation, we will also
recall the origin and the relativistic value of the kink
stationary speed used in this work. The bias current and
dissipation present in the system have an opposite effect on
fluxon motion leading to mutual equilibration at a certain speed
\cite{Scott1979}. The dynamics of the soliton in the homogenous
system is described by the equation
\begin{equation}\label{40}
 \partial_t^2 \phi + \alpha \, \partial_t \phi-
\partial_x^2 \phi +  \sin \phi = -\Gamma_0 .
\end{equation}
If we multiply both sides of this equation by the time derivative
of the field  $\phi$ and next integrate it with respect to the
space variable, then we obtain
\begin{equation}\label{41}
\frac{d}{d t} H^{SG}= - \int_{-\infty}^{+\infty} d x \left[
\Gamma_0 \,
\partial_t \phi + \alpha \, (\partial_t \phi)^2 \right] ,
\end{equation}
where $H^{SG}$ is the hamiltonian of the sine-Gordon model
$$
H^{SG} = \int_{-\infty}^{+\infty} dx  \left[ \frac{1}{2}
(\partial_t \phi)^2 + \frac{1}{2} (\partial_x \phi)^2 +  (1-\cos
\phi) \right] .
$$
Introducing the kink ansatz
$$
\phi(t,x) = 4 \arctan \left( \frac{x-x_0 - u
t}{\sqrt{1-u^2}}\right)
$$
into equation (42) leads to the ordinary differential equation for
the fluxon velocity
\begin{equation}\label{42}
\frac{d u}{d t} = \frac{1}{4} \, \pi \Gamma_0 (1 -
u^{2})^{\frac{3}{2}} - \alpha \, u (1 - u^2) .
\end{equation}
The constant equilibrium $(du/dt=0)$ solution of this equation,
corresponds to the situation when the power input caused by the
bias current is is balanced by the loss of power due to
dissipation
\begin{equation}\label{43}
u_s = \frac{1}{\sqrt{1+(\frac{4 \alpha}{\pi \Gamma_0})^2}} ~.
\end{equation}
This velocity describes the stationary motion of the fluxon in the
homogeneous junction.


\begin{thebibliography}{99}

\bibitem{Josephson1962}
B. D. Josephson, Phys. Lett. \textbf{1}, 251 (1962).

\bibitem{Josephson1974}
B. D. Josephson, Rev. Mod. Phys. \textbf{46}, 251 (1974).

\bibitem{Anderson1963}.
P. W. Anderson, J. M. Rowell,  Phys. Rev. Lett. \textbf{10}, 230
(1963).

\bibitem{Seidel2015}
Applied Superconductivity: Handbook on Devices and Applications,
P. Seidel (ed.), Wiley, ISBN 13: 978-3-27-741209-9 (2015).

\bibitem{Braginski2019}
A. I. Braginski,  J Supercond. Nov. Magn. \textbf{32}, 23 (2019).

\bibitem{Bednorz1986}
J. G. Bednorz and K. A. M\"{u}ller, Z. Phys. B \textbf{64}, 189
(1986).

\bibitem{Schilling1993}
A. Schilling, M. Cantoni, J. D. Guo, and H. R. Ott, Nature
\textbf{363}, 56 (1993).

\bibitem{Benabdallah1996}
A. Benabdallah, J. G. Caputo, and A. C. Scott, Phys. Rev. B
\textbf{54}, 16139 (1996).

\bibitem{Kemp2002a}
A. Kemp, A. Wallraff, and A. V. Ustinov, Phys. Stat. Sol. B
\textbf{233}, 472 (2002).

\bibitem{Kemp2002b}
A. Kemp, A. Wallraff, and A.V. Ustinov, Physica C \textbf{368},
324 (2002).

\bibitem{Gulevich2006a}
D. R. Gulevich and F. V. Kusmartsev, Phys. Rev. Lett. \textbf{97},
017004 (2006).

\bibitem{Gulevich2008}
D. R. Gulevich, F. V. Kusmartsev, S. Savelev, V. A. Yampol\'skii,
and F. Nori, Phys. Rev. Lett. \textbf{101}, 127002 (2008).

\bibitem{Caputo2014}
J.G. Caputo and D. Dutykh, Phys. Rev. E \textbf{90}, 022912
(2014).

\bibitem{Monaco2016}
R. Monaco, J. Phys. Condens. Matter \textbf{28}, 445702 (2016).

\bibitem{Monaco2018}
R. Monaco, J. Mygind, and V. P. Koshelets, Supercond. Sci.
Technol. \textbf{31}, 025003 (2018).

\bibitem{Dobrowolski2020}
T. Dobrowolski and A. Jarmoli\'nski, Phys. Rev. E \textbf{101},
052215 (2020).

\bibitem{Dobrowolski2009}
T. Dobrowolski, Phys. Rev. E \textbf{79}, 046601 (2009).

\bibitem{Dobrowolski2011}
T. Dobrowolski, Discrete Contin. Dyn. Syst. Ser. S \textbf{4},
1095 (2011).

\bibitem{Dobrowolski2012}
T. Dobrowolski, Ann. Phys. (N.Y.) \textbf{327}, 1336 (2012).

\bibitem{Dobrowolski2013}
T. Dobrowolski, Eur. Phys. J. B \textbf{86}, 346 (2013).

\bibitem{Dobrowolski2017}
T. Dobrowolski and A. Jarmoli\'nski, Phys. Rev. E \textbf{96},
012214 (2017).

\bibitem{Gatlik2021}
J. Gatlik and T. Dobrowolski, Physica D \textbf{428}, 133061
(2021).

\bibitem{Fei1992}
Z. Fei, Y. S. Kivshar, and L. Vazquez, Phys. Rev. A \textbf{46},
5214 (1992).

\bibitem{Campbell1983}
D. K. Campbell, J. F. Schonfeld, and C.A. Wingate, Physica D
\textbf{9}, 1 (1983).

\bibitem{Goodman2005}
R.H. Goodman and R. Haberman, SIAM J. Appl. Dyn. Syst.
\textbf{4},  1195 (2005).

\bibitem{Goodman2008}
R.H. Goodman, Chaos \textbf{18}, 023113 (2008).

\bibitem{Manton2021}
N. S. Manton, K. Ole\'s, T. Roma\'nczukiewicz, and A.
Wereszczy\'nski Phys. Rev. D \textbf{103}, 025024 (2021).

\bibitem{Adam2022}
C. Adam, N. S. Manton, K. Ole\'s, T. Roma\'nczukiewicz, and A.
Wereszczy\'nski, Phys.Rev. D \textbf{105}, 065012 (2022).

\bibitem{Takyi2016}
I. Takyi and H. Weigel, Phys. Rev. D \textbf{94}, 085008 (2016).

\bibitem{Kevrekidis2019}
P. G. Kevrekidis and R. H. Goodman, arXiv:1909.03128v1 (2019).

\bibitem{Dorey2011}
P. Dorey, K. Mersh, T. Roma\'nczukiewicz, and Y. Shnir, Phys. Rev.
Lett. \textbf{107}, 091602 (2011).

\bibitem{Scott1979}
D.W. McLaughlin and A.C. Scott,  Phys. Rev. A \textbf{18}, 1652
(1978).





\end{thebibliography}
\end{document}